\documentclass[letterpaper,prb,twocolumn,showpacs,superscriptaddress]{revtex4}
\usepackage[latin1]{inputenc}
\usepackage{ifpdf,ae}
\usepackage{amsmath}
\usepackage{amsfonts}
\usepackage{amssymb}

\begin{document}

\title{Magnetization damping in a local-density approximation}
\author{Hans Joakim Skadsem} \affiliation{Centre for Advanced Study at
  the Norwegian Academy of Science and Letters, Drammensveien 78,
  NO-0271 Oslo, Norway} \affiliation{Department of Physics, Norwegian
  University of Science and Technology, N-7491 Trondheim, Norway}
\author{Yaroslav Tserkovnyak} \affiliation{Centre for Advanced Study
  at the Norwegian Academy of Science and Letters, Drammensveien 78,
  NO-0271 Oslo, Norway} \affiliation{Department of Physics and
  Astronomy, University of California, Los Angeles, California 90095,
  USA} \author{Arne Brataas} \affiliation{Centre for Advanced Study at
  the Norwegian Academy of Science and Letters, Drammensveien 78,
  NO-0271 Oslo, Norway} \affiliation{Department of Physics, Norwegian
  University of Science and Technology, N-7491 Trondheim, Norway}
\author{Gerrit E. W. Bauer} \affiliation{Kavli Institute of
  NanoScience, Delft University of Technology, Lorentzweg 1, 2628 CJ
  Delft, The Netherlands} \date{\today}


\begin{abstract}
  The linear response of itinerant transition metal ferromagnets to
  transverse magnetic fields is studied in a self-consistent adiabatic
  local-density approximation. The susceptibility is calculated from a
  microscopic Hamiltonian, including spin-conserving impurities,
  impurity induced spin-orbit interaction and magnetic impurities
  using the Keldysh formalism. The Gilbert damping constant in the
  Landau-Lifshitz-Gilbert equation is identified, parametrized by an
  effective transverse spin dephasing rate, and is found to be
  inversely proportional to the exchange splitting. Our result justify
  the phenomenological treatment of transverse spin dephasing in the
  study of current-induced magnetization dynamics in weak, itinerant
  ferromagnets by Tserkovnyak~\textit{et~al.}.
  \cite{tserkovnyak05:_curren} We show that neglect of gradient
  corrections in the quasiclassical transport equations leads to
  incorrect results when the exchange potential becomes of the order
  of the Fermi energy.
\end{abstract}
\maketitle

\section{Introduction}

\label{sec:introduction}

The drive to miniaturize and reduce power demands of electronic
appliances motivates research in nanoscale magnetoelectronics,
\textit{i.e.} the science and technology that exploits additional
functionalities offered by ferromagnets integrated into electronic
circuits and devices. Spectacular advances have been realized already
in the last decade, mainly in the area of magnetic disk and magnetic
random access memories. Itinerant transition metals and its alloys are
the materials of choice for magnetoelectronic applications due to
their high electric conductivity and Curie temperatures. Increasing
speed and reducing energy demands of switching a bit of information
encoded by the magnetization direction of a ferromagnetic grain is one
of the key problems in the field. A thorough understanding of the
dynamics of the magnetization order parameter in transition metals is
necessary to make progress in this direction.

Phenomenologically, the low temperature magnetization dynamics in
ferromagnets is well described by the Landau-Lifshitz-Gilbert (LLG)
equation.\cite{lifshitz80:_statis_physic_part,gilbert} Ferromagnetic
resonance (FMR) experiments can be fitted to obtain accurate values
for the parameters of the LLG equation, \textit{viz.} the Gilbert
constant that parametrizes viscous damping and the effective
(demagnetization and crystal anisotropy) fields. The LLG phenomenology
has been successfully applied to explain a rich variety of dynamic
magnetic phenomena.\cite{bhagat74:_temper,heinrich93:_ultrat} Some
progress has been made in predicting magnetic crystal anisotropies by
first-principles calculations.\cite{daalderop90:_first} However, in
spite of being a crucial device parameter that governs the switching
time of magnetic memory elements, the material dependence of the
intrinsic magnetization damping has not yet been understood. The
Gilbert damping parameter also plays an important role in
current-induced magnetization excitations and domain wall
motion.\cite{zhang04:_roles,tserkovnyak05:_curren}

Deriving a microscopic description of the dynamics of transition metal
ferromagnets is a formidable task; even the nature of the ground state
is still under debate. Two different viewpoints can be distinguished.
On one hand, ferromagnetism can be seen to be caused by the atomic
correlations in partially filled and essentially localized $d$
orbitals. The $s$-electrons that are responsible for electron
transport are in this picture affected by the magnetic order only
indirectly via local exchange interactions. Such physics is expressed
by the so-called $s$-$d$ model in which the spin of localized
$d$-electrons, $\boldsymbol{S}_{i}$, interact with free $s$-electron
spins, $\boldsymbol{s}(\boldsymbol{r}_{i},t)$, through a Heisenberg
exchange term.

In the opposite point of view, the $d$-electrons are not only
broadened into bands, but are also strongly hybridized with
neighboring $s$-$p$ orbitals. A separate treatment of the orbital
symmetry is then not warranted for the description of low energy
properties at long time scales. The Stoner model represents the
essence of this itinerant magnetism in terms of two (minority and
majority) parabolic energy bands that are split by a constant exchange
potential. Spin density-functional theory in a local density
approximation is the modern version of itinerant magnetism, forming
the basis of most band structure calculations to date. The nature of
the real wave function of \textit{3d} ferromagnets that combines
features of both extremes is presumably captured by sophisticated
many-body frameworks such as the dynamical mean-field model. It is at
present not obvious, however, how to compute the low-energy collective
dynamics of ferromagnets taking disorder as a well as local
correlations into account.

The Gilbert damping coefficient in the LLG equation, usually denoted
by $\alpha$, has attracted quite some theoretical attention.
Incoherent scattering of electron-hole pair excitations by phonons and
magnons is a possible mechanism by which energy and angular momentum
can be dissipated. Heinrich~\textit{et~al.}\cite{heinrich67} suggested
a model in which conduction electron spins become polarized by
scattering with magnons. The spin angular momentum is subsequently
transferred to the lattice by spin-orbit mediated relaxations. The
resulting damping coefficient was found to be proportional to the
electronic scattering rate, $\alpha\sim\tau^{-1}$. We will return to
this result in Sec.~\ref{sec:homog-ferr}. More recently, a
phenomenological treatment of the Gilbert damping has also been
reported in Ref.~\onlinecite{piechon06:_theor_i}.

A different relaxation process was proposed in
Ref.~\onlinecite{kambersky70:_landau_lifsh}, and further elaborated in
Refs.~\onlinecite{korenman72:_anomal,korenman74:_impur,kunes02:_first,kunes03:_errat}:
In the presence of spin-orbit interaction, the electronic energy
levels depend on a time-dependent magnetization direction, giving rise
to the notion of a \textquotedblleft breathing Fermi
surface\textquotedblright. The time lag of the electronic distribution
response to a moving magnetization vector is equivalent with
dissipation. In this model the Gilbert damping coefficient is
proportional to the scattering time, $\alpha\sim\tau$. Extrinsic
contributions to the FMR linewidth such as eddy currents excited by
time-dependent magnetic fields,\cite{ament55:_elect} sample
inhomogeneities, or two-magnon scattering processes
\cite{sparks61:_ferrom_relax,lutovinov79:_relax,mills03:_spin_dynam_confin_magnet_struc}
have been suggested as well.

In diluted magnetic semiconductors such as (Ga,Mn)As the magnetism
originates mainly from the local spins of the half-filled spin-$5/2$
Mn $d$-shells. The spins are coupled by a local exchange interaction
to the valence band holes, and non-locally, via the holes,
ferromagnetically to each other. The holes contribute only slightly to
the magnetization, but are exclusively responsible for the finite
conductivity. The $s$-$d$ model is therefore appropriate for
understanding the magnetization damping in ferromagnetic
semiconductors.\cite{sinova04:_magnet_ga_mn_as,tserkovnyak04:_mean}
Magnetization damping in the $s$-$d$ model can be understood in terms
of the so-called spin-pumping mechanism.\cite{spinpumping,
  mills03:_ferrom,simanek03:_gilber,simanek03a:_gilber} The motion of
the localized spins pumps a spin current into the conduction electron
bath, in which the thus created spin accumulation is dissipated by
spin-flip scattering. Ref.~\onlinecite{tserkovnyak04:_mean} reported a
non-monotonous dependence of the damping on the scattering rate,
\textit{i.e}. $\alpha \sim \tau^{-1}$ for clean and $\alpha\sim \tau$
for dirty samples. As mentioned above, the $s$-$d$ model does not
necessarily give a good description of transport and dynamical
properties of transition metal ferromagnets.  The notion of
$d$-electrons pumping spins into an $s$-electron system becomes
doubtful when the hybridization is very strong. Recently it has been
demonstrated that the magnetization dynamics in the $s$-$d$ model and
in an itinerant Stoner model can be quite different
indeed.\cite{tserkovnyak05:_curren} For example, for a given spin-flip
relaxation mechanism, the Gilbert damping is significantly suppressed
in the $s$-$d$ description by a factor of the (usually small) fraction
of the total magnetization carried by the delocalized $s$-electrons.

Recently Kohno~\textit{et~al.}\cite{kohno06:_micro} reported results
for both the Gilbert damping and the non-adiabatic current-induced
spin torque term postulated by Zhang and Lee\cite{zhang04:_roles}
which is parameterized by a constant $\beta$.  The results in
Ref.~\onlinecite{tserkovnyak05:_curren} have been obtained under the
assumption that the exchange splitting is small compared to the Fermi
energy. The diagrammatic treatment by Kohno~\textit{et~al.} was not
restricted to weak ferromagnets, but the differences turned out to be
very small for transition metals.\cite{tserkovnyak05:_curren}

In the present paper we generalize the treatment of the transverse
spin dephasing of Ref.~\onlinecite{tserkovnyak05:_curren} beyond the
relaxation time approximation. We relax the previous limitation to
weak ferromagnets and derive the corresponding Gilbert damping. We use
a self-consistent adiabatic local-density approximation (ALDA) model
in the presence of a dilute concentration of scalar and magnetic
impurities, as well as spin-orbit interaction originating from
impurities, and we demonstrate how to generalize the previous
treatment to strong ferromagnets. The generalization of the Keldysh
approach is non-trivial and introduces subtle, but important gradient
corrections. 

Our main result is that for spatially homogeneous itinerant
ferromagnets, the Gilbert damping constant is given by
\begin{equation}
  \alpha = \frac{\hbar}{\Delta\tau_{\perp}},
\end{equation}
where $\Delta$ is the modulus of the local-density
exchange-correlation potential, and $\tau_{\perp}^{-1}$ is a
transverse (Bloch) spin dephasing rate caused by spin-orbit
interaction and magnetic disorder. This appears to be at variance with
Kohno~\textit{et~al.},\cite{kohno06:_micro} who find a Gilbert damping
constant that depends on both the longitudinal and the transverse
scattering rates. Except for this issue, we obtain the same detailed
expression for $\alpha$. This is gratifying since these theoretical
machineries are completely different.

This paper is organized in the following way: The microscopic model,
as well as the simplifying ALDA are presented in
Sec.~\ref{sec:adiab-local-dens}, while the linear response formalism,
microscopically and phenomenologically, will be treated in
Sec.~\ref{sec:linearresponse}. The detailed derivation of the linear
response function, starting from the Keldysh Green function formalism,
is the topic of Sec.~\ref{sec:micr-deriv-susc}. Conclusions are
summarized in Sec.~\ref{sec:conclusion}.

\section{Time-dependent adiabatic local-density approximation (ALDA)}

\label{sec:adiab-local-dens}

Density-functional theory (DFT) is a successful and widely used method
to study electronic structure and magnetism in transition metal
ferromagnets.\cite{kubler00:_theor} In the Kohn-Sham implementation,
non-interacting pseudo-particles are introduced that exhibit the same
ground state density as the interacting many-electron system.  This is
realized by introducing a fictitious exchange-correlation potential
that has to be determined self-consistently by energy
minimization.\cite{kohn99:_nobel_lectur} DFT can be expanded to handle
time-dependent phenomena in systems out of
equilibrium.\cite{runge84:_densit}

We study the magnetization dynamics in a simplified time-dependent DFT
in the local spin-density approximation, in which non-interacting
Kohn-Sham particles are treated as free electrons. We realize that the
transition metals have complex energy bands and wave functions also in
the local density approximation and that even integrated properties
such as conductivities are not necessarily well described by free
electrons.  However, since we are interested in dirty systems, in
which an additional scrambling occurs by elastic impurity scattering
between different bands, we are confident that our treatment is a good
starting point for more sophisticated computations that take the full
band structure into account.

Our model for itinerant ferromagnetism is described by the Hamiltonian
\begin{multline}
  \label{eq:hamiltoniansdft}
  \hat{\mathcal{H}} = \hat{1} \left[ \mathcal{H}_{0} +
    V(\boldsymbol{r})\right] + \frac{1}{2} \gamma \hbar
  H\hat{\sigma}_{z} + \frac{1}{2} \gamma \hbar
  \boldsymbol{H}_{\mathrm{xc}} [\hat{\rho}(\boldsymbol{r},t)] \cdot
  \hat{\boldsymbol{\sigma}} \\ + \frac{1}{2} \gamma \hbar
  \boldsymbol{h}(\boldsymbol{r},t) \cdot \hat{\boldsymbol{\sigma}} +
  \hat{V}_{\mathrm{so}}(\boldsymbol{r}) +
  \hat{V}_{\mathrm{m}}(\boldsymbol{r} ).
\end{multline}
Matrices in $2\times2$ spin space are denoted by a hat ($\hat
{\phantom{\sigma}}$). Spin-independence is indicated by the unit
matrix $\hat{1}$, and $\hat{\boldsymbol{\sigma}}$ is a vector of the
Pauli matrices. Here, $\mathcal{H}_{0}$ is the
translationally-invariant Hamiltonian of non-interacting electrons,
and $V(\boldsymbol{r})$ is the elastic spin-conserving impurity
potential. $H>0$ is an effective magnetic field in the $z$-direction
consisting of internal anisotropy fields and externally applied
contributions, and $-\gamma<0$ is the electronic gyromagnetic ratio.
Electron-electron interactions are described by the
exchange-correlation (vector) field $\boldsymbol{H}_{\mathrm{xc}}$.
The weak, transverse magnetic driving field is denoted
$\boldsymbol{h}(\boldsymbol{r},t)$, and the potentials due to impurity
induced spin-orbit interaction and a magnetic disorder configuration
are denoted $\hat{V}_{\mathrm{so}}(\boldsymbol{r})$ and
$\hat{V}_{\mathrm{m}}(\boldsymbol{r})$, respectively.

The published approximation schemes for time-dependent
exchange-correlation functionals are still rather crude and/or
untested. The simplest approximation is known as the adiabatic
local-density approximation (ALDA). Here the conventional LDA
exchange-correlation potential is adopted for the instantaneous
time-dependent density.\cite{zangwill80:_reson,gross85:_local} The
ALDA potential is therefore local in both spatial and temporal degrees
of freedom and reduces for the current problem to
\begin{equation}
  \label{eq:alda}
  \gamma \hbar \boldsymbol{H}_{\mathrm{xc}} [\hat{\rho}]
  (\boldsymbol{r},t) \approx \Delta \boldsymbol{m}(\boldsymbol{r},t),
\end{equation}
where $\Delta$ is an effective exchange splitting constant and
$\boldsymbol{m}(\boldsymbol{r},t)$ is the local magnetization
direction of the ferromagnet.  By construction, the exchange field is
always parallel to the magnetization direction, and thus automatically
satisfies the zero-torque theorem.\cite{capelle01:_spin} Since we are
interested in the low-energy transverse magnetization dynamics the
(atomic-scale) position dependence of $\Delta$ is disregarded.

In the ALDA, the exchange-correlation potential is released from a
possible functional dependence on the history of the system. The ALDA
should therefore be valid only when the system is close to the
equilibrium configuration, \textit{i.e.} for a slowly varying
magnetization direction in both space and time. This is the case when
$\hbar\partial_{t} \ll \Delta$, and $\partial_{\boldsymbol{r}} \ll
k_{F}$, with $k_{F}$ being a characteristic Fermi wave vector.
Improved descriptions of the exchange potential have been proposed
(\textit{e.g.} the generalized gradient approximation), but for slowly
varying uniform perturbations, such corrections are believed to be
small.\cite{qian02:_spin}

\section{Linear response}
\label{sec:linearresponse}

For a weak magnetic driving field, the response of the ferromagnet can
be formulated within linear response theory. An expression for the
response to the perturbative field $\boldsymbol{h}(\boldsymbol{r},t)$
is derived quantum mechanically from the ALDA Hamiltonian defined by
Eqs.~(\ref{eq:hamiltoniansdft},~\ref{eq:alda}) in
Sec.~\ref{sec:quant-line-resp}. The response derived from the
phenomenological Landau-Lifshitz-Gilbert equation is presented in
Sec.~\ref{sec:land-lifsh-gilb}. These results are then used in
Sec.~\ref{sec:micr-deriv-susc} to find a microscopic expression for
the Gilbert damping coefficient.

\subsection{Quantum linear response}
\label{sec:quant-line-resp}

The Kubo formalism provides expressions for the linear response to a
time-dependent perturbation. The response functions can be derived by
considering the time evolution of the non-equilibrium density matrix.
Starting from the effective Hamiltonian~(\ref{eq:hamiltoniansdft}) in
the ALDA of Eq.~(\ref{eq:alda}), the small time-dependent perturbation
operator should include the self-consistent exchange as:
\begin{displaymath}
  H_{\mathrm{int}}(t) = \int\mathrm{d}\boldsymbol{r}\: \left[
    \frac{\Delta}{\hbar} \delta \boldsymbol{m}(\boldsymbol{r},t) +
    \gamma\boldsymbol{h} (\boldsymbol{r},t) \right]
  \cdot\boldsymbol{s}(\boldsymbol{r}), 
\end{displaymath}
where $\boldsymbol{s}(\boldsymbol{r})$ is the spin density operator.
The emphasis of this paper is on the transverse, non-equilibrium
components of the spin density. We denote
$s_{0}=|\boldsymbol{s}_{0}|$, $\langle\boldsymbol{s} \rangle= - s_{0}
\boldsymbol{e}_{z} + \langle \delta \boldsymbol{s} \rangle$, where
$\langle\delta\boldsymbol{s} \rangle \perp \boldsymbol{e}_{z}$. Hence,
$|\boldsymbol{s}| = s_{0}$, and $\boldsymbol{m} = - \langle
\boldsymbol{s} \rangle/s_{0}$ in the ALDA.

For axially symmetric systems, the non-equilibrium spin density
response can be expressed conveniently in terms of $\delta s_{\pm} =
\delta s_{x} \pm\mathrm{i} \delta s_{y}$. The transverse part of the
response to the magnetic field can then be written as
\begin{equation}
  \label{eq:quantumlinearresponse}
  \langle\delta s_{-} (\boldsymbol{q},\omega)
  \rangle= - \chi_{-+}(\boldsymbol{q},\omega) \left[
    \frac{\Delta}{\gamma\hbar}\delta m_{-} (\boldsymbol{q},\omega) +
    h_{-} (\boldsymbol{q},\omega) \right],
\end{equation}
where the retarded susceptibility tensor
\begin{displaymath}
  \chi_{\mu\nu}(\boldsymbol{r},\boldsymbol{r}^{\prime};t) =
  \frac{\mathrm{i} \gamma}{2\hbar}\Theta(t) \langle[
  s_{\mu}(\boldsymbol{r},t), s_{\nu}(\boldsymbol{r}^{\prime},0)]
  \rangle, 
\end{displaymath}
has been introduced. The brackets $[\dots]$ indicate a commutator, and
the angular brackets $\langle\dots\rangle$ a thermodynamical average.
In the derivation of the above expression, we have made use of axial
symmetry under which $\chi_{++} = \chi_{--} = 0$. In the ALDA,
Eq.~(\ref{eq:quantumlinearresponse}) can be simplified to
\begin{equation}
  \label{eq:quantumlinearresponseLDA}
  \langle\delta s_{-} (\boldsymbol{q},\omega) \rangle = -
  \tilde{\chi}_{-+}(\boldsymbol{q},\omega) h_{-}
  (\boldsymbol{q},\omega), 
\end{equation}
where the self-consistent linear response to the driving field
\begin{equation}
  \label{eq:transform}
  \tilde{\chi}_{-+}^{-1}(\boldsymbol{q},\omega) =
  \chi_{-+}^{-1}(\boldsymbol{q},\omega) - \frac{\Delta}{\gamma\hbar
    s_{0}}. 
\end{equation}
has been introduced. Hence, in the ALDA, the linear response of an
interacting system reduces itself to calculating the response
$\chi_{-+}$ of a noninteracting system with a fixed (Stoner
enhancement) exchange field.\cite{green93:_spin}

\subsection{Landau-Lifshitz-Gilbert susceptibility}

\label{sec:land-lifsh-gilb}

The phenomenological Landau-Lifshitz equation
\cite{lifshitz80:_statis_physic_part} is widely used to model
transverse magnetization dynamics. The magnetization direction
$\boldsymbol{m}(\boldsymbol{r},t)$ is treated as a classical field
whose dynamics is governed by an effective magnetic field
$\boldsymbol{H}_{\mathrm{eff}}(\boldsymbol{r},t)$, obtainable from the
free-energy functional of the system, $F[\boldsymbol{M}]$:
\begin{displaymath}
  \boldsymbol{H}_{\mathrm{eff}}(\boldsymbol{r},t) =
  -\partial_{\boldsymbol{M}}F[\boldsymbol{M}].
\end{displaymath}
The Landau-Lifshitz equation describes undamped (\textit{i.e.},
free-energy conserving) precessional motion about the local effective
magnetic field:
\begin{displaymath}
  \partial_{t} \boldsymbol{m}(\boldsymbol{r},t) = - \gamma
  \boldsymbol{m}(\boldsymbol{r},t)
  \times\boldsymbol{H}_{\mathrm{eff}}(\boldsymbol{r},t),
\end{displaymath}
preserving the magnitude of the magnetization. The field
$\boldsymbol{H}_{\mathrm{eff}}(\boldsymbol{r},t)$ includes
contributions from external, exchange, demagnetization and
crystal-anisotropy magnetic fields.

The Landau-Lifshitz equation does not dissipate energy since the
effective magnetic field always points normal to the instantaneous
constant-free-energy surfaces. However, the electronic degrees of
freedom do not respond infinitely fast to the magnetization dynamics,
which means that in reality the effective field is a functional of the
time-dependent magnetization at previous times. A finite lag in the
response of the dynamics corresponds to energy dissipation.  In a
magnetic system, the energy-loss implies a lowering of the Zeeman
energy by a torque in the direction of the cross product of
magnetization and its time derivative; the energy-loss can be
parametrized by the phenomenological Gilbert damping constant
$\alpha$.\cite{gilbert} Hence, we arrive at the
Landau-Lifshitz-Gilbert (LLG) equation:
\begin{equation*}
  \partial_{t} \boldsymbol{m} = -\gamma\boldsymbol{m} \times
  \boldsymbol{H}_{\mathrm{eff}} + \alpha\boldsymbol{m} \times
  \partial_{t} \boldsymbol{m}. 
\end{equation*}
Here, $\boldsymbol{H}_{\mathrm{eff}}$ only depends on the
instantaneous magnetic configuration of the ferromagnet. Generally,
the damping is a tensor quantity with symmetries reflecting the
crystal structure,\cite{mills03:_spin_dynam_confin_magnet_struc} but
in practice anisotropic corrections to damping are small compared to
those in the free-energy. \cite{heinrich88:_struc_ni_fe_ag}

Assuming an external field of the form
$\gamma\boldsymbol{H}_{\mathrm{eff}}(\boldsymbol{r},t) =
\omega_{0}(\boldsymbol{r},t) \boldsymbol{e}_{z}$, and a small rf
driving field $\boldsymbol{h}(\boldsymbol{r},t)$, the excited
small-angle transverse magnetization dynamics can be computed easily
by the linearized LLG equation,
\begin{displaymath}
  m_{-} (\boldsymbol{q},\omega) = \frac{\gamma h_{-}
    (\boldsymbol{q},\omega )}{\omega_{0}(\boldsymbol{q},\omega) -
    \omega - \mathrm{i}\alpha(\boldsymbol{q},\omega) \omega}.
\end{displaymath}
which corresponds to a susceptibility
\begin{displaymath}
  \tilde{\chi}_{-+}(\boldsymbol{q},\omega) = \frac{\gamma
    s_{0}}{\omega_{0}(\boldsymbol{q},\omega) - \omega -
    \mathrm{i} \alpha(\boldsymbol{q},\omega) \omega},
\end{displaymath}
that can be directly compared with the microscopic response function
$\chi_{-+}$ by Eq.~(\ref{eq:transform}). Assuming that $\Delta/ \hbar
\gg (\omega,\omega_{0})$ one
obtains\cite{sinova04:_magnet_ga_mn_as,tserkovnyak04:_mean}
\begin{equation}
  \label{eq:alphasusceptibility}
  \alpha(\boldsymbol{q},\omega \to 0) =
  \frac{\Delta^{2}}{\gamma\hbar^{2} s_{0}} \lim_{\omega \to 0}
  \partial_{\omega}\mathrm{Im} \chi_{-+}(\boldsymbol{q},\omega).
\end{equation}
Hence, finding a microscopic expression for the Gilbert damping is
equivalent to determining the quantum mechanical transverse
susceptibility.

It is worth noticing that, in general, the damping coefficient may
depend on the spin-wave wave vector $\boldsymbol{q}$. A damping of the
form $\alpha(\boldsymbol{q},\omega\rightarrow0)\sim q^{2}$ will
introduce an additional dissipative term in the LLG equation,
$\partial_{t} \boldsymbol{m} \propto - \alpha \boldsymbol{m} \times
\nabla^{2} \partial_{t} \boldsymbol{m}$, which is similar in form to
the exchange field $\propto \boldsymbol{m} \times \nabla^{2}
\boldsymbol{m}$ for anisotropic ferromagnets. The main emphasis in
this paper, however, is on the isotropic part of the Gilbert damping.

\section{Microscopic derivation of the susceptibility}

\label{sec:micr-deriv-susc}

In this section we determine the susceptibility function according to
time-dependent spin DFT in the ALDA for a disordered ferromagnet. The
Keldysh Green function formalism is used, and the assumption of weak
and slowly varying perturbations in space and time allows us to use
the simplifying gradient expansion. Finally,
Eq.~(\ref{eq:alphasusceptibility}) is invoked to obtain the Gilbert
damping coefficient.

\subsection{Kinetic equation}
\label{sec:gener-boltzm-equat}

We proceed from the ALDA Hamiltonian (see
Sec.~\ref{sec:adiab-local-dens}):
\begin{multline}
  \hat{\mathcal{H}} = \hat{1} \left[ \mathcal{H}_{0} +
    V(\boldsymbol{r}) \right] + \frac{1}{2} (\Delta + \gamma \hbar H)
  \hat{\sigma}_{z} \\ + \frac{1}{2} \gamma \hbar
  \boldsymbol{h}(\boldsymbol{r},t)\cdot \hat{\boldsymbol{\sigma}} +
  \hat{V}_{\mathrm{so}}(\boldsymbol{r}) +
  \hat{V}_{\mathrm{m}}(\boldsymbol{r}).
\end{multline}
In the following discussion, we assume a homogeneous static magnetic
field $H$, define $\Delta^{\prime} = \Delta + \gamma \hbar H$, and
drop the prime for brevity.

The impurities are assumed to be randomly distributed over positions
$\boldsymbol{r}_{i}$ with short-range, scalar disorder potentials
\begin{displaymath}
  V(\boldsymbol{r}) = \sum_{i} v_{0}(\boldsymbol{r}_{i}) \delta
  (\boldsymbol{r} - \boldsymbol{r}_{i}).
\end{displaymath}
The scattering potentials are Gaussian-distributed with zero average
and a white noise correlator
\begin{displaymath}
  \langle V(\boldsymbol{r}) V(\boldsymbol{r}^{\prime}) \rangle = \xi
  \delta (\boldsymbol{r} - \boldsymbol{r}^{\prime}).
\end{displaymath}
We define a characteristic scattering time $\tau$ by $\xi^{-1} = \pi
(\nu_{\uparrow} + \nu_{\downarrow}) \tau / \hbar$, with $\nu_{s}$
being the density of states at the Fermi level for electrons with spin
$s$. The spin-orbit interaction associated to impurities is described
by the potential
\begin{displaymath}
  \hat{V}_{\mathrm{so}} (\boldsymbol{r}) = i \beta
  \hat{\boldsymbol{\sigma}} \cdot (\boldsymbol{\nabla}
  V(\boldsymbol{r}) \times \boldsymbol{\nabla}),
\end{displaymath}
where $\beta$ is a spin-orbit interaction strength given by $- \hbar^2
/ 4m_e^2 c^2$, in terms of the electron mass $m_e$ and the speed of
light $c$. The magnetic disorder in the ferromagnet is modeled as
\begin{displaymath}
  \hat{V}_{\mathrm{m}}(\boldsymbol{r}) = \sum_{i} v_{\mathrm{m}}
  (\boldsymbol{r}_{i}) \delta (\boldsymbol{r} - \boldsymbol{r}_{i})
  \boldsymbol{\mathcal{S}}(\boldsymbol{r}_{i}) \cdot
  \hat{\boldsymbol{\sigma}}, 
\end{displaymath}
where $\boldsymbol{\mathcal{S}}(\boldsymbol{r}_{i})$ denotes the spin
of an impurity at position $\boldsymbol{r}_{i}$. The internal degrees
of freedom of the magnetic impurities are assumed to be frozen. Also
the vector impurity exchange field $\boldsymbol{V}(\boldsymbol{r}) =
\frac{1}{2} \mathrm{Tr} [ \hat{V}_{\mathrm{m}} (\boldsymbol{r})
\hat{\boldsymbol{\sigma}}]$ is taken to be distributed according to
Gaussian white noise characteristics, \textit{i.e.}  $\langle
V_{\alpha} (\boldsymbol{r}) \rangle = 0$ and
\begin{displaymath}
  \left \langle V_{\alpha}(\boldsymbol{r}) V_{\beta}
    (\boldsymbol{r}^{\prime})  \right \rangle
  =\xi_{m}^{(\alpha)} \delta_{\alpha\beta} \delta( \boldsymbol{r} -
  \boldsymbol{r}^{\prime}), 
\end{displaymath}
where $\alpha,\beta$ denote spatial components of the vector field,
and the strength of the second moment is given by
\begin{displaymath}
  \xi_{m}^{(\alpha)} = \left \{
    \begin{array}
      [c]{ll}
      \xi_{\perp}, & \alpha = x, y\\
      \xi_{||}, & \alpha = z,
    \end{array}
  \right.
\end{displaymath}
similar to Ref.~\onlinecite{kohno06:_micro}.

We employ the Keldysh Green function formalism \cite{keldysh} to
calculate the spin susceptibility defined earlier. This method has
distinct advantages over the equilibrium formalism when it comes to
describing non-equilibrium phenomena, but also some drawbacks.
However, as we will see in Sec.~\ref{sec:homog-ferr} in particular,
whereas the equilibrium formalism requires tedious calculations of
vertex corrections, the Keldysh formalism requires that one carefully
accounts for subtle gradient corrections in order to obtain the
correct dynamics. 

The Green function in Keldysh space, denoted by an inverted caret
($\check{\phantom{\sigma}}$) takes on the form\cite{rammersmith}
\begin{displaymath}
  \check{G}(1,2) =
  \begin{pmatrix}
    \hat{G}^{R}(1,2) & \hat{G}^{K}(1,2)\\
    0 & \hat{G}^{A}(1,2)
  \end{pmatrix}.
\end{displaymath}
The retarded, advanced and Keldysh Green functions are given by
\begin{align*}
  \hat{G}^{R}(1,2) & = -\mathrm{i} \Theta(t_{1} - t_{2}) \langle\{
  \Psi (1),\Psi^{\dagger} (2) \} \rangle,\\
  \hat{G}^{A}(1,2) & = +\mathrm{i}\Theta(t_{2} - t_{1})
  \langle\{\Psi(1) , \Psi^{\dagger}(2) \} \rangle,
\end{align*}
and
\begin{displaymath}
  \hat{G}^{K}(1,2) = -\mathrm{i}\langle[\Psi(1),\Psi^{\dagger}(2)]
  \rangle,
\end{displaymath}
respectively. Brackets $[\dots]$ indicate a commutator, the curly
brackets $\{\dots\}$ an anti-commutator, while $\Psi^{(\dagger)}$ is a
fermion annihilation (creation) operator. In this notation, all field
variables (position, time and spin) are contained in the numerical
indices $1$ and $2$.

In the presence of slowly varying perturbations, the two-point
propagator variables can be transformed into the Wigner
representation, \textit{viz.} the center-of-mass coordinates and the
Fourier transform of the Green function with respect to the relative
coordinates:
\begin{displaymath}
  \check{G}(X,k) = \int\mathrm{d}x \:  \mathrm{e}^{-\mathrm{i}k x}
  \check{G}(X + x/2, X - x/2).
\end{displaymath}
Here, a four-vector formulation has been introduced, where the vector
for the center-of-mass coordinates is $X = (\boldsymbol{R},T)$, the
corresponding relative coordinates are given by $x =
(\boldsymbol{r},t)$, and finally $k = (\boldsymbol{k},\varepsilon)$.
The four-vector product is defined as $k\cdot x = -\varepsilon t +
\boldsymbol{k} \cdot \boldsymbol{r}$. The Wigner representation is
particularly convenient when the variation of the Green function on
center coordinates is slow on the scale of the Fermi wavelength, since
this allows us to perform a gradient expansion in these coordinates.
Subtracting the Dyson equation and its conjugate, one finds the
relation
\begin{equation}
  \label{eq:dyson}
  [\check{G}^{-1}_{0} - \check{\Sigma} \stackrel{\otimes}{,}
  \check{G}] = 0,
\end{equation}
where the symbol $\otimes$ denotes a convolution (in position, time
and spin), the commutator corresponds to the $2\times2$ Keldysh matrix
structure, and $\check{G}^{-1}_{0}$ is the inverse of the
retarded/advanced Green function in absence of any impurities. It is
diagonal in Keldysh space, and in the Wigner representation has the
structure
\begin{equation}
  \label{eq:17}
  \hat{G}^{-1}_{0}(\boldsymbol{R},T;\boldsymbol{k},\varepsilon) =
  \hat{1}(\varepsilon- \varepsilon_{k}) - \frac{1}{2}
  \Delta\hat{\sigma}_{z} - \frac{1}{2} \gamma
  \hbar\boldsymbol{h}(\boldsymbol{R},T) \cdot
  \hat{\boldsymbol{\sigma}} 
\end{equation}
in spin space, with $\varepsilon_{k}$ denoting the free-electron
energy measured with respect to the chemical potential. The final
component in Eq.~\eqref{eq:dyson}, $\check{\Sigma}$, is the
self-energy due to the impurity configurations and spin-orbit
interaction. One can show by a formal Taylor expansion that the
convolution can be represented by
\begin{equation}
  \label{eq:starproduct}
  (A\otimes B)(X,k) = \mathrm{e}^{\mathrm{i} (\partial_{X}^{A}
    \cdot\partial_{k}^{B} - \partial_{k}^{A} \cdot
    \partial_{X}^{B})/2} A(X,k) B(X,k)
\end{equation}
in the Wigner representation.\cite{rammersmith}

Physical quantities such as occupation probabilities and densities are
expressible in terms of the distribution Green function $\hat{G}^{<}$,
which is given by the combination
\begin{displaymath}
  \hat{G}^{<}=\frac{1}{2}\left( \hat{G}^{K} + \mathrm{i} \hat{A}
  \right),
\end{displaymath}
where we have introduced the spectral function $\hat{A} =
\mathrm{i}(\hat{G}^{R} - \hat{G}^{A})$. To derive a kinetic equation
for $\hat{G}^{<}$, we subtract the diagonal components of
Eq.~\eqref{eq:dyson}, and combine the result with the Keldysh
component of the same equation. In summary, one finds the kinetic
equation
\begin{equation*}
  \left[ \hat{G}^{R}\right]^{-1} \otimes \hat{G}^{<} - \hat{G}^{<}
  \otimes \left[ \hat{G}^{A} \right]^{-1} = \hat{\Sigma}^{<} \otimes
  \hat{G}^{A} - \hat{G}^{R} \otimes \hat{\Sigma}^{<}. 
\end{equation*}
Assuming slowly varying perturbations, we now use the gradient
expansion, in which the exponential in Eq.~(\ref{eq:starproduct}) is
expanded, and only the first two terms of the expansion are
kept.\cite{kadanoff} This results in a simplified kinetic equation for
the distribution Green function, \textit{viz.}
\begin{multline}
  \label{eq:kineticglessergradient}
  \lbrack\hat{G}_{0}^{-1} , \hat{G}^{<}] + \frac{\mathrm{i}}{2}
  [\hat{G}_{0}^{-1} , \hat{G}^{<}]_{p} -
  \frac{\mathrm{i}}{2} [\hat{G}^{<}, \hat{G}_{0}^{-1}]_{p} \\
  - (\hat{\Sigma}^{R} \hat{G}^{<} - \hat{G}^{<} \hat{\Sigma}^{A}) +
  (\hat{G}^{R} \hat{\Sigma}^{<} - \hat{\Sigma}^{<} \hat{G}^{A})\\
  = \frac{\mathrm{i}}{2} \left([ \hat{\Sigma}^{R}, \hat{G}^{<}]_{p} -
    [\hat{G}^{<},\hat{\Sigma}^{A}]_{p} \right)\\
  - \frac{\mathrm{i}}{2} \left([ \hat{G}^{R},\hat{\Sigma}^{<}]_{p} -
    [\hat{\Sigma }^{<} , \hat{G}^{A}]_{p}\right).
\end{multline}
All terms to first order in the generalized Poisson bracket are kept,
$[\hat{X},\hat{Y}]_{p} = \partial_{X} \hat{X} \cdot \partial_{k}
\hat{Y} - \partial_{k} \hat{X} \cdot \partial_{X} \hat{Y}$, where the
four-vector notation implies $\partial_{X} \cdot \partial_{k} =
\partial_{\boldsymbol{R}} \cdot \partial_{\boldsymbol{k}} - \hbar
\partial_{T} \partial_{\varepsilon}$. We see that the gradient
expansion reduces the complex convolution of the Dyson
equation~(\ref{eq:dyson}) to the $2\times2$ matrix multiplication of
Eq.~(\ref{eq:kineticglessergradient}).

It is important to carefully include all contributions to first order
in the Poisson brackets. The correct dynamics is captured only when
gradient corrections to the spectral function are kept, and, in the
case of anisotropically distributed magnetic impurities, those to the
self-energies as well. Such gradient corrections are caused by the
non-uniform driving field. For simple systems, \textit{e.g.} normal
metals, such corrections are absent. {\em This means that one should
  not expect naive generalizations of Boltzmann equations to strong
  ferromagnets to be completely correct.}

The final ingredient that transforms
Eq.~(\ref{eq:kineticglessergradient}) into a useful kinetic equation
is an expression for the self-energy. For weak impurity scattering,
the self-consistent Born approximation is appropriate:
\begin{displaymath}
  \check{\Sigma}(1,2) = \langle \hat{V}_{\mathrm{tot}} (1)
  \check{G}(1,2) \hat{V}_{\mathrm{tot}}(2)\rangle,
\end{displaymath}
where $\hat{V}_{\mathrm{tot}}$ is the total potential (which is
diagonal in Keldysh space), angular brackets $\langle \dots \rangle$
denotes impurity potential averaging, while $\check{G}(1,2)$ is
already an impurity averaged Green function. This expression can be
separated into four different self-energy contributions. The
spin-conserving impurity scattering is described by
\begin{displaymath}
  \check{\Sigma}_{\mathrm{imp}}(\boldsymbol{R},T; \boldsymbol{k},
  \varepsilon) = \xi \int
  \frac{\mathrm{d}\boldsymbol{k}^{\prime}}{(2\pi)^{3}} \:
  \check{G}(\boldsymbol{R},T;\boldsymbol{k}^{\prime},\varepsilon). 
\end{displaymath}
Introducing $\boldsymbol{n} = \boldsymbol{k} \times
\boldsymbol{k}^{\prime}$, the terms arising from the spin-orbit
interaction can be written as
\begin{multline*}
  \check{\Sigma}_{\mathrm{so}}^{(1)} (\boldsymbol{R},T;\boldsymbol{k},
  \varepsilon) = -\mathrm{i} \xi \beta
  \int\frac{\mathrm{d}\boldsymbol{k}^{\prime}}{(2\pi)^3} \big[
  \check{G}(\boldsymbol{R},T;\boldsymbol{k}^{\prime},\varepsilon)
  \hat{\boldsymbol{\sigma}} \cdot \boldsymbol{n} \\ - \boldsymbol{n}
  \cdot \hat{\boldsymbol{\sigma}}
  \check{G}(\boldsymbol{R},T;\boldsymbol{k}^{\prime},\varepsilon)
  \big],
\end{multline*}
and
\begin{displaymath}
  \check{\Sigma}_{\mathrm{so}}^{(2)}
  (\boldsymbol{R},T;\boldsymbol{k},\varepsilon) = \xi \beta^2 \int
  \frac{\mathrm{d}\boldsymbol{k}^{\prime}}{(2\pi)^3} 
  \boldsymbol{n}\cdot\hat{\boldsymbol{\sigma}}
  \check{G}(\boldsymbol{R},T;\boldsymbol{k}^{\prime},\varepsilon) 
  \hat{\boldsymbol{\sigma}} \cdot \boldsymbol{n}.
\end{displaymath}
The magnetic impurity configuration results in
\begin{displaymath}
  \check{\Sigma}_{\mathrm{m}}(\boldsymbol{R},T; \boldsymbol{k},
  \varepsilon) = \sum_{i=x,y,z} \xi_{m}^{(i)}
  \int\frac{\mathrm{d}\boldsymbol{k}^{\prime}}{(2\pi)^{3}}\:
  \hat{\sigma}_{i}\check{G}(\boldsymbol{R} ,
  T;\boldsymbol{k}^{\prime} , \varepsilon)\hat{\sigma}_{i}. 
\end{displaymath}
Finally, to make connection with the spin density, one can use that
\begin{equation*}
  \langle\boldsymbol{s}(\boldsymbol{R},T)\rangle =
  \frac{\hbar}{4\mathrm{i}\pi}\int_{-\infty}^{\infty}
  \mathrm{d}\varepsilon
  \int\frac{\mathrm{d} \boldsymbol{k}}{(2\pi)^{3}} \mathrm{Tr} \left\{
    \hat{\boldsymbol{\sigma}}
    \hat{G}^{<}(\boldsymbol{R},T;\boldsymbol{k}, \varepsilon)
  \right\}.
\end{equation*}
All necessary quantities are now defined, and a kinetic equation for
the distribution Green function can be derived. In the next section,
the details are worked out for a bulk, single-domain ferromagnet.

\subsection{Homogeneous ferromagnet}
\label{sec:homog-ferr}

We concentrate in the following on the Gilbert damping constant in the
limit of vanishing spin-wave wave vector, $\boldsymbol{q}\rightarrow0$
as measured in FMR experiments. In this limit (and the ALDA), only
spin-orbit interaction and magnetic disorder can transfer angular
momentum out of the spin dynamics into the lattice.  Without it, spin
and orbital degrees of freedom are completely decoupled and the
Gilbert constant vanishes. Spin waves with finite wave lengths may
decay also by spin-conserving scattering, which is likely to dominate
magnetic impurity or spin-orbit interaction scattering when
$q$ becomes larger.\cite{tserkovnyak05:_spin}

We simplify the notation by defining the time and energy-dependent
density matrix
\begin{equation*}
  \hat{\rho}(T,\varepsilon) = \int
  \frac{\mathrm{d}\boldsymbol{k}}{(2\pi)^{3}}
  \hat{G}^{<}(T;\boldsymbol{k},\varepsilon), 
\end{equation*}
and solve Eq.~\eqref{eq:kineticglessergradient} to obtain a diffusion
equation for this quantity. In detail, we find that
\begin{widetext}
  \begin{align}
    \nonumber & \mathrm{i}\hbar\partial_T \hat{\rho} - \frac{1}{2}
    \Delta[ \hat{\sigma}_z,\hat{\rho}] - \frac{1}{2}\gamma \hbar
    \boldsymbol{h}(T)\cdot [\hat{\boldsymbol{\sigma}},\hat{\rho}] +
    \frac{\mathrm{i}}{4}\gamma \hbar^2 \{ \partial_T
    (\boldsymbol{h}\cdot\hat{\boldsymbol{\sigma}}) ,
    \partial_{\varepsilon}\hat{\rho}\} \\ \nonumber = & \sum_{i}
    \xi_m^{(i)} \int\frac{\mathrm{d}\boldsymbol{k}}{(2\pi)^3} \left(
      \hat{\sigma}_{i} \hat{G}^R \hat{\sigma}_{i} \hat{\rho} -
      \hat{G}^R \hat{\sigma}_{i} \hat{\rho}\hat{\sigma}_{i} - \left(
        \hat{\rho}\hat{\sigma}_{i} \hat{G}^A \hat{\sigma}_{i} -
        \hat{\sigma}_{i}
        \hat{\rho}\hat{\sigma}_{i} \hat{G}^A \right) \right) \\
    + & \sum_{i} \frac{\mathrm{i}\xi_m^{(i)}}{2}
    \int\frac{\mathrm{d}\boldsymbol{k}}{(2\pi)^3} \left(
      [\hat{\sigma}_{i} \hat{G}^R \hat{\sigma}_{i}, \hat{\rho}]_p -
      [\hat{G}^R, \hat{\sigma}_{i}\hat{\rho}\hat{\sigma}_{i}]_p -
      \left( [\hat{\rho},\hat{\sigma}_{i} \hat{G}^A
        \hat{\sigma}_{i}]_p -
        [\hat{\sigma}_{i}\hat{\rho}\hat{\sigma}_{i}, \hat{G}^A]_p
      \right) \right) \\ \nonumber + & \sum_{i,j} \xi\beta^2
    \int\frac{\mathrm{d}\boldsymbol{k}^{\prime}}{(2\pi)^3}
    \int\frac{\mathrm{d}\boldsymbol{k}}{(2\pi)^3} n_i n_j \left(
      \hat{\sigma}_i \hat{G}^R \hat{\sigma}_j \hat{G}^< - \hat{G}^R
      \hat{\sigma}_i \hat{G}^< \hat{\sigma}_j - \left( \hat{G}^<
        \hat{\sigma}_i \hat{G}^A \hat{\sigma}_j - \hat{\sigma}_i
        \hat{G}^< \hat{\sigma}_j \hat{G}^A \right) \right) \\
    \nonumber + & \sum_{i,j} \frac{\xi\beta^2}{2}
    \int\frac{\mathrm{d}\boldsymbol{k}^{\prime}}{(2\pi)^3}
    \int\frac{\mathrm{d}\boldsymbol{k}}{(2\pi)^3} n_i n_j \left(
      [\hat{\sigma}_i \hat{G}^R \hat{\sigma}_j, \hat{G}^<]_p -
      [\hat{G}^R, \hat{\sigma}_i \hat{G}^< \hat{\sigma}_j]_p - \left(
        [\hat{G}^<, \hat{\sigma}_i \hat{G}^A \hat{\sigma}_j]_p -
        [\hat{\sigma}_i \hat{G}^< \hat{\sigma}_j, \hat{G}^A]_p \right)
    \right).
  \end{align}
\end{widetext}
Here, the arguments to $\hat{\rho}(T,\varepsilon$,
$\hat{G}^{R/A}(T;k,\varepsilon)$ and $\hat{G}^<(T;k^{\prime},
\varepsilon)$ are not written out explicitly for the sake of notation.
Summation indices $i,j$ run over Cartesian components $x,y,z$. On the
left hand side of the equation, we recognize precession around the
fixed exchange field and the driving field, as well as a gradient term
due to the non-uniformity of the driving field. On the right hand side
we find collision integrals due to spin-orbit interaction and magnetic
impurities. We also find gradient corrections to the collision
integrals in the third and fifth line above. These corrections are
often neglected, but are important for strong ferromagnets, to be
discussed below. As explained, scalar disorder does not affect the
uniform spin dynamics and drop out of the kinetic equation.

For the response function $\chi_{-+}(\omega)$ introduced in
Sec.~\ref{sec:quant-line-resp}, we need to find an expression for
$\langle\delta s_{-}(\omega)\rangle$, the transverse part of the spin
density.  To this end, we extract the upper right matrix component of
$\hat{\rho}(T,\varepsilon)$, a matrix component we simply denote
$\delta s_{-}(T,\varepsilon)$. This is now related to the
non-equilibrium spin density by
\begin{displaymath}
  \langle\delta s_{-}(T)\rangle = \frac{\hbar}{2\mathrm{i}\pi}
  \int_{-\infty}^{\infty} \mathrm{d}\varepsilon \: \delta
  s_{-}(T,\varepsilon). 
\end{displaymath}
With $\nu(\varepsilon_{k})$ being the density of states at energy
$\varepsilon_{k}$, we find that
\begin{widetext}
  \begin{align}
    \nonumber & \mathrm{i}\hbar\partial_T \delta s_- (T,\varepsilon) -
    \Delta \delta s_-(T,\varepsilon) + \gamma \hbar h_-(T)
    \rho_z(\varepsilon) + \frac{\mathrm{i}}{2}\gamma \hbar^2
    \partial_T h_- (T) \int \mathrm{d}\varepsilon_k \:
    \nu(\varepsilon_k)
    \partial_{\varepsilon} G^<_d (k,\varepsilon) \\ \nonumber = &
    -2\mathrm{i}(\xi_{\perp} + \xi_{||}) \int \mathrm{d}\varepsilon_k
    \: \nu(\varepsilon_k) \left[ A_d(k,\varepsilon) \delta
      s_-(T,\varepsilon) - \rho_d(\varepsilon) A_- (T;k,\varepsilon)
    \right] \\ \nonumber & - \hbar (\xi_{\perp} - \xi_{||})
    \int\mathrm{d}\varepsilon_k \: \nu(\varepsilon_k) \left[
      \partial_{\varepsilon} A_z (k,\varepsilon) \partial_T \delta
      s_-(T,\varepsilon) - \partial_{\varepsilon} \rho_z(\varepsilon)
      \partial_T A_- (T;k,\varepsilon) \right] \\
    & - 4(\xi_{\perp} - \xi_{||}) \int\mathrm{d}\varepsilon_k \:
    \nu(\varepsilon_k) \left[ \mathrm{Re}G^R_z(k,\varepsilon) \delta
      s_-(T,\varepsilon) + \rho_z(\varepsilon) \mathrm{Re}G^R_-
      (T;k,\varepsilon) \right] \\ \nonumber & + 2\mathrm{i}\hbar
    (\xi_{\perp} + \xi_{||}) \int \mathrm{d}\varepsilon_k \:
    \nu(\varepsilon_k) \left[
      \partial_{\varepsilon} \rho_d(\varepsilon)\partial_T
      \mathrm{Re}G^R_- (T;k,\varepsilon) + \partial_{\varepsilon}
      \mathrm{Re}G^R_d(k,\varepsilon) \partial_T\delta
      s_-(T,\varepsilon) \right] \\ \nonumber & -
    \frac{8\mathrm{i}}{9} \xi\beta^2 \int\mathrm{d}\varepsilon_k \:
    \nu(\varepsilon_k)k^2 \int\mathrm{d}\varepsilon_{k^{\prime}}\:
    \nu(\varepsilon_{k^{\prime}}) k^{\prime 2} \left[
      A_d(k,\varepsilon)
      G^<_- (T;k^{\prime},\varepsilon) - G^<_d
      (k^{\prime},\varepsilon) A_- (T;k,\varepsilon) \right] \\
    \nonumber & + \frac{8\mathrm{i}\hbar}{9} \xi \beta^2
    \int\mathrm{d}\varepsilon_k \: \nu(\varepsilon_k) k^2
    \int\mathrm{d}\varepsilon_{k^{\prime}} \:
    \nu(\varepsilon_{k^{\prime}}) k^{\prime 2} \left[
      \partial_{\varepsilon} G^<_d(k^{\prime},\varepsilon) \partial_T
      \mathrm{Re}G^R_- (T;k,\varepsilon) +
      \partial_{\varepsilon} \mathrm{Re}G^R_d(k,\varepsilon)
      \partial_T G^<_- (T;k^{\prime},\varepsilon)\right],
  \end{align}
where have used the convenient matrix notation
\begin{equation}
  \label{eq:4}
  \hat{G} = \hat{1}G_d +
  \boldsymbol{G}\cdot\hat{\boldsymbol{\sigma}}.
\end{equation}
We now need to Fourier transform and integrate this formidable
equation over energy to obtain a diffusion equation for $\langle
\delta s_- (\omega) \rangle$. Before we proceed with this calculation,
we notice that the real part of the response function simply
determines the resonance condition for the system and is thus
unimportant for the determination of the Gilbert damping. With this in
mind, we write
\begin{align}
  \label{eq:9}
  \nonumber (\hbar\omega - \Delta) \langle \delta & s_- (\omega)
  \rangle = \frac{\hbar}{\pi}(\xi_{\perp} + \xi_{||})
  \int_{-\infty}^{\infty}\mathrm{d}\varepsilon
  \int\mathrm{d}\varepsilon_k \: \nu(\varepsilon_k) \left[
    \rho_d(\varepsilon) A_- (\omega;k,\varepsilon) -
    A_d(k,\varepsilon) \delta s_- (\omega,\varepsilon) \right] \\ + &
  \frac{\hbar^2 \omega}{2\pi} (\xi_{\perp} - \xi_{||})
  \int_{-\infty}^{\infty} \mathrm{d}\varepsilon
  \int\mathrm{d}\varepsilon_k \: \nu(\varepsilon_k) \left[
    \partial_{\varepsilon} A_z(k,\varepsilon) \delta s_-
    (\omega,\varepsilon) -
    \partial_{\varepsilon} \rho_z(\varepsilon)
    A_-(\omega;k,\varepsilon) \right] \\ \nonumber + &
  \frac{4\hbar}{9\pi} \xi\beta^2 \int_{-\infty}^{\infty}
  \mathrm{d}\varepsilon \: \int\mathrm{d}\varepsilon_k \:
  \nu(\varepsilon_k) k^2 \int\mathrm{d}\varepsilon_{k^{\prime}} \:
  \nu(\varepsilon_{k^{\prime}}) k^{\prime 2} \left[ G^<_d
    (k^{\prime},\varepsilon) A_- (\omega;k,\varepsilon) -
    A_d(k,\varepsilon) G^<_- (\omega;k^{\prime},\varepsilon) \right] +
  \mathcal{F}(\omega)h_- (\omega),
  \end{align}
\end{widetext}
where $\mathcal{F}(\omega)$ is real, and thus does not contribute to
the damping. The dissipation is now determined by the above integrals
over energy and momentum. Notice how the signs between the
longitudinal and transverse magnetic impurity scattering strength
enter in the above equation. For simple isotropic magnetic impurities,
\textit{i.e.} with $\xi_{\perp}=\xi_{||}$, the second line of
Eq.~\eqref{eq:9} does not contribute to the dissipative dynamics.
This term is due to gradient corrections involving self-energies in
the original kinetic equation \eqref{eq:kineticglessergradient}. To
correctly capture the dynamics when the magnetic impurities are
anisotropically distributed, it is essential to include such gradient
corrections as well.

In order to calculate the integrals, we need expressions for $G^<_-$,
$\delta s_{-}$ and the spectral function $\hat{A}$.  Since these
quantities enter the collision integrals, we can solve for $\delta
s_{-}$ and $\hat{A}$ to zeroth order in scattering rates. To this end,
we solve Eq.~(\ref{eq:kineticglessergradient}) for the Fourier
transform of $G^<_-$, and find that
\begin{multline}
  \label{eq:8}
  G^<_{-}(\omega;k^{\prime},\varepsilon) \approx \frac{\gamma\hbar
    h_{-} (\omega)}{\Delta} \left( 1 + \frac{\hbar\omega}{\Delta}
  \right) G^<_{z}(k^{\prime},\varepsilon) \\ + \frac{\gamma\hbar^{2}
    \omega h_{-} (\omega)}{2\Delta} \partial_{\varepsilon} G^{<}_{d}
  (k^{\prime},\varepsilon) +
  \mathcal{O}\left(\xi,\xi_m^{(i)}\right).
\end{multline}
At this point we take $G^{<}_{d} (k^{\prime},\varepsilon) =
\mathrm{i}n_{F}(\varepsilon) A_{d}(k^{\prime},\varepsilon)$, with
$n_{F}$ the Fermi-Dirac distribution, so that
\begin{displaymath}
  \partial_{\varepsilon} G^{<}_{d}(k^{\prime},\varepsilon) =
  -\mathrm{i}\delta (\varepsilon) A_{d}(k^{\prime},\varepsilon) +
  \mathrm{i}n_{F}(\varepsilon)
  \partial_{\varepsilon}A_{d}(k^{\prime},\varepsilon)
\end{displaymath}
at low temperatures. Additionally, we use that
\begin{displaymath}
  \delta s_- (\omega,\varepsilon) =
  \int\frac{\mathrm{d}\boldsymbol{k}^{\prime}}{(2\pi)^3} G^<_- (\omega;
  k^{\prime},\varepsilon),
\end{displaymath}
in combination with Eq.~(\ref{eq:8}) to solve for the second quantity
in question.

In the dilute limit, the Lorentzian shape of the spectral function
approaches a Dirac delta function, and two quasiparticle spin bands,
split by the exchange field $\Delta$, are resolved. For a uniform,
time-independent transverse magnetic field one finds
\begin{equation}
  \label{eq:specnograds}
  \hat{A}_{0}(k,\varepsilon) = \pi\sum_{s} \delta
  (\varepsilon - \varepsilon_{ks}) \left( \hat{1} + s\hat{\sigma}_{z} + s
    \frac{\gamma\hbar\boldsymbol{h} \cdot
      \hat{\boldsymbol{\sigma}}}{\Delta}\right).
\end{equation}
The two spin bands are denoted by $s = \uparrow,\downarrow= \pm$, and
the notation $\varepsilon_{ks} = \varepsilon_{k} + s\Delta/2$ has been
introduced.  A non-uniform driving will also introduce terms in the
spectral function that are linear in gradients. A detailed derivation
of the spectral function to first order in Poisson brackets for a
time-dependent driving field $\boldsymbol{h}(T)$ is given in
App.~\ref{sec:spectral-function}, with the main result being
\begin{multline}
  \hat{A}(T;k,\varepsilon) = \hat{A}_{0}(T;k,\varepsilon) \\
  + \frac{\mathrm{i}\gamma\hbar^{2}}{\Delta^{2}} \left( A_{z}
    (k,\varepsilon) + \frac{\Delta}{2} \partial_{\varepsilon}
    A_{d}(k,\varepsilon)\right) \hat{\sigma}_{z} \partial_{T} \left(
    \boldsymbol{h}\cdot\hat {\boldsymbol{\sigma}}\right) ,
\end{multline}
where $\hat{A}_{0}$ is the spectral function from
Eq.~\eqref{eq:specnograds}.  We see that the weak, transverse driving
field induce off-diagonal gradient corrections to the
``instantaneous'' spectral function. The diagonal components are
unchanged, and are given by
\begin{displaymath}
  A_{d}(k,\varepsilon) = \pi\sum_{s} \delta(\varepsilon-
  \varepsilon_{ks}),
\end{displaymath}
and
\begin{displaymath}
  A_{z}(k,\varepsilon) = \pi\sum_{s} s \delta(\varepsilon-
  \varepsilon_{ks}). 
\end{displaymath}

We are now in a position to calculate the above energy integrals. To
be more specific, considering the integrals due to magnetic disorder,
we find that these terms become
\begin{multline*}
  \int\mathrm{d}\varepsilon_{k} \: \nu(\varepsilon_{k}) \left[
    \rho_{d}(\varepsilon) A_{-}(\omega;k,\varepsilon) -
    A_{d}(k,\varepsilon)
    \delta s_{-}(\omega,\varepsilon) \right] \\
  = \frac{\mathrm{i}\pi^{2} \gamma\hbar^{2} h_{-} (\omega)
    \omega}{2\Delta} \delta(\varepsilon) \left( \nu_{\uparrow} +
    \nu_{\downarrow}\right) ^{2},
\end{multline*}
and
\begin{multline*}
  \hbar\omega\int\mathrm{d}\varepsilon_{k} \: \nu(\varepsilon_{k})
  \left[ \partial_{\varepsilon} A_{z}(k,\varepsilon) \delta s_{-}
    (\omega,\varepsilon) - \partial_{\varepsilon}\rho_{z}(\varepsilon)
    A_{-}(\omega;k,\varepsilon) \right] \\
  = \frac{\mathrm{i}\pi^{2}\gamma\hbar^{2}
    h_{-}(\omega)\omega}{\Delta} \delta(\varepsilon) (\nu_{\uparrow} -
  \nu_{\downarrow})^{2}.
\end{multline*}
One can derive analogous results also for the spin-orbit contribution
in Eq.~(\ref{eq:9}); inserted we find that
\begin{multline*}
  (\hbar\omega- \Delta) \langle\delta s_{-}(\omega) \rangle
  \approx\mathcal{F}(\omega) h_{-} \\
  + \frac{2\mathrm{i}\pi\gamma \hbar^3}{9\Delta} h_- \omega \xi
  \beta^2 \left[ (\nu_{\uparrow}^2 k_{F\uparrow}^4 +
    \nu_{\downarrow}^2 k_{F\downarrow}^4) +
    2\nu_{\uparrow}\nu_{\downarrow} k_{F\uparrow}^2 k_{F\downarrow}^2
  \right] \\ + \frac{\mathrm{i}\pi\gamma\hbar^{3}}{\Delta} h_{-}
  \omega\left[ \xi_{\perp} (\nu_{\uparrow}^{2} + \nu_{\downarrow}^{2})
    + 2\xi_{||} \nu_{\uparrow}\nu_{\downarrow} \right].
\end{multline*}
Using Eq.~\eqref{eq:quantumlinearresponseLDA} to identify (the low
frequency) $\mathrm{Im}\chi_{-+}$, we find from
Eq.~\eqref{eq:alphasusceptibility} that the Gilbert damping is given
by
\begin{multline}
  \label{eq:13}
  \alpha = \frac{2\pi\hbar}{9s_0} \xi\beta^2 \left[ (\nu_{\uparrow}^2
    k_{F\uparrow}^4 + \nu_{\downarrow}^2 k_{F\downarrow}^4) +
    2\nu_{\uparrow}\nu_{\downarrow} k_{F\uparrow}^2 k_{F\downarrow}^2
  \right] \\ + \frac{\pi\hbar}{s_{0}} \left[\xi_{\perp}
    (\nu_{\uparrow}^{2} + \nu_{\downarrow}^{2}) +
    2\xi_{||}\nu_{\uparrow} \nu_{\downarrow} \right],
\end{multline}
which agrees with the diagrammatic calculation of
Kohno~\textit{et~al.}.\cite{kohno06:_micro}

Next, we would like to relate the Gilbert damping constant in
Eq.~\eqref{eq:13} to other physical quantities. Comparing a
Bloch-Bloembergen\cite{tserkovnyak05:_curren,bloembergen50} equation
of motion for the magnetization vector with the corresponding LLG
equation, we find for weak driving fields and small angle
magnetization dynamics that
\begin{equation}
  \label{eq:14}
  \alpha=\frac{\hbar}{\Delta\tau_{\perp}},
\end{equation}
where, in our case,
\begin{equation*}
  \frac{1}{\tau_{\perp}}= \frac{1}{\tau_{\mathrm{so}}} +
  \frac{1}{\tau_{\mathrm{m}}}.
\end{equation*}
Above we have defined the effective transverse scattering rates from
spin-orbit interaction and magnetic impurities, \textit{viz.}
\begin{equation*}
  \frac{1}{\tau_{\mathrm{so}}} = \frac{2\pi\Delta}{9s_0} \xi\beta^2
  \left[(\nu_{\uparrow}^2 k_{F\uparrow}^4 + \nu_{\downarrow}^2
    k_{F\downarrow}^4) + 2\nu_{\uparrow}\nu_{\downarrow}
    k_{F\uparrow}^2 k_{F\downarrow}^2 \right], 
\end{equation*}
and
\begin{equation*}
  \frac{1}{\tau_{\mathrm{m}}} = \frac{\pi\Delta}{s_{0}}\left[
    \xi_{\perp}
    (\nu_{\uparrow}^{2}+\nu_{\downarrow}^{2}) + 2\xi_{||}\nu_{\uparrow} 
    \nu_{\downarrow}\right].
\end{equation*}
By comparison, the longitudinal spin relaxation rate obtained from
\textit{e.g.} Fermi's Golden rule reads
\begin{equation*}
  \frac{1}{\tau_{||}}=\frac{4\pi}{\hbar}( \nu_{\uparrow} +
  \nu_{\downarrow}) \left[ \xi_{||} + \frac{2}{9} \xi\beta^2
    k_{F\uparrow}^2 k_{F\downarrow}^2 \right]. 
\end{equation*}
For weak ferromagnets, the density of states and momentum at the Fermi
energy is not strongly spin dependent, \textit{i.e.}
$\nu_{s}\simeq\nu_{F}$, and $k_{Fs} \simeq k_F$. Therefore $2s_{0}
\approx\hbar\Delta\nu_{F}$, which implies equal transverse and
longitudinal scattering rates for impurity induced spin-orbit
interaction and \textit{isotropic} magnetic impurity scattering,
\textit{i.e.} $\tau_{\perp} = \tau_{||}$.\cite{kohno06:_micro}

Since we succeeded in reproducing the general diagrammatic result of
Ref.~\onlinecite{kohno06:_micro}, we also identified the necessary
measure to transcend the semiclassical treatment of
Ref.~\onlinecite{tserkovnyak05:_curren}. Most important are the
gradient corrections to the spectral function, but in the presence of
anisotropically distributed magnetic impurities, gradient corrections
to the self-energies should be included as well.

In a metal, the longitudinal spin-orbit induced scattering time
depends on the spin-conserving elastic scattering time.
Experimentally, one typically finds that the ratio of spin-conserving
to non-spin-conserving scattering events
$\epsilon=\tau/\tau_{\parallel}$ is not very sensitive to the
concentration of impurities.\cite{spinpumping} This means that in
systems where spin-orbit induced dephasing dominates the Gilbert
damping, $\alpha$ is proportional to the resistivity of the system,
\begin{equation*}
  \alpha=\frac{\hbar ne^{2}}{2\Delta m}\rho\epsilon,
\end{equation*}
where $n$ is the electron concentration, and $\rho$ is the
resistivity. A linear relation between Gilbert damping and resistivity
was found in a recent experimental study of electronic transport in
thin permalloy films by
Ingvarsson~\textit{et~al.}\cite{ingvarsson02:_role_ni_fe}

The present discussion is mainly focused on the low temperature
regime, but the linear dependence on the scattering rates suggests
that the damping should increase with increasing temperatures. It is
experimentally known that $\epsilon$ is not very sensitive to
variations in temperature.\cite{jedema03:_spin} Consequently, we
expect that the Gilbert damping constant is proportional to the
resistivity also at higher temperatures.

Since the Gilbert damping coefficient is proportional to
$\Delta^{-1}$, dissipation is reduced in the strong ferromagnet limit.
Since the damping in Eq.~(\ref{eq:14}) depends on the transverse spin
dephasing rate, Gilbert damping does not vanish in half metals in
which the chemical potential falls below the band edge of one of the
spin bands.

The form of the damping coefficient found in
Eqs.~(\ref{eq:13},~\ref{eq:14}) agrees with previous studies on
spin-flip mechanisms for magnetization damping. Nearly four decades
ago, Heinrich~\textit{et~al.}\cite{heinrich67} suggested, based on the
$s$-$d$ exchange interaction between localized $d$ electrons and
itinerant $s$ electrons, that electron-hole pairs could be excited by
magnons. Assuming that the exchange splitting is much larger than the
spin-flip rate, and denoting the fraction of the total spin carried by
the delocalized electrons by $\eta< 1$, they found $\alpha=
\eta\hbar^{2} \nu _{F}/2s_{0} \tau_{\mathrm{sf}}$ in the long wave
length limit. The result is expressed in terms of
$\tau_{\mathrm{sf}}$, a phenomenological electron-hole pair life time.
This result can be compared with Eqs.~(\ref{eq:13},~\ref{eq:14}) by
using the approximation $2s_{0} \approx\hbar\Delta\nu_{F}$.  We see
that when the magnetization is mainly carried by the $d$-electrons,
which are not affected by spin-flip scattering, the predicted damping
in the $s$-$d$ model is much weaker than in our Stoner model.

\section{Conclusion}

\label{sec:conclusion}

In conclusion, we present a kinetic equation for the distribution
matrix of itinerant ferromagnets in the adiabatic local-density
approximation. The spin susceptibility and Gilbert damping constant is
obtained microscopically for a homogeneous ferromagnet by the Keldysh
Green function formalism. Magnetization damping arises from magnetic
disorder in the ferromagnet, and we have shown that it is important to
keep all terms to linear order in Poisson brackets to obtain the
correct result in presence of impurity induced spin-orbit interaction
magnetic disorder. The Gilbert coefficient can be expressed in terms
of an effective transverse spin dephasing rate that has been
introduced earlier as a phenomenological constant
\cite{tserkovnyak05:_curren}. Our framework could also be generalized
to handle first-principles band structure calculations for specific
types of impurities and disorder, that may be the starting point for
systematic studies of the Gilbert damping as a function of material
parameters.

\begin{acknowledgements}
  This work was supported by the Research Council of Norway through
  Grants No. 162742/V00, 158518/431, 158547/431, and 167498/V30, and
  by EC Contract IST-033749 DynaMax.
\end{acknowledgements}

\appendix

\section{Spectral function}
\label{sec:spectral-function}

In this section we derive the spectral function to first order in
Poisson brackets, in the presence of a weak, time-dependent transverse
driving field $\boldsymbol{h}(T)$. Since the spectral functions
appearing in Eq.~\eqref{eq:9} are already proportional to scattering
rates, we will not keep gradient terms involving self energies in the
following derivation.

To proceed, we consider $\hat{G}^{R}$, which is determined from the
relation
\begin{displaymath}
  \hat{G}^{R} \otimes\left[  \hat{G}^{R} \right] ^{-1} = \hat{1},
\end{displaymath}
where $[\hat{G}^{R}]^{-1} = \hat{G}^{-1}_{0} - \hat{\Sigma}^{R}$, and
$\hat {G}^{-1}_{0}$ is the inverse Green function in the clean limit,
\textit{viz.}  Eq.~\eqref{eq:17}. To first order in Poisson brackets,
one can show that in the Wigner representation
\begin{displaymath}
  \hat{G}^{R}(T;\boldsymbol{k},\varepsilon) = \left(  \hat{1} -
    \frac{\mathrm{i}}{2} \left[\hat{G}^{R}_{0}, \left(
        \hat{G}^{R}\right)^{-1}\right]_{p} \right)  \hat{G}^{R}_{0} +
  \mathcal{O} ([\dots]_{p}^{2}),
\end{displaymath}
where now
\begin{multline*}
  \hat{G}^{R}_{0} (T;\boldsymbol{k},\varepsilon) =\\
  \frac{1}{\mathrm{det} \left[ (\hat{G}^{R})^{-1} \right] }
  \begin{pmatrix}
    \scriptstyle \varepsilon- \varepsilon_{k} + \frac{1}{2}\Delta-
    \Sigma^{R}_{22} & \scriptstyle \frac{1}{2}\gamma\hbar h_{-} +
    \Sigma^{R}_{-}\\[0.2cm] \scriptstyle \frac{1}{2}\gamma\hbar h_{+}
    + \Sigma^{R}_{+} & \scriptstyle \varepsilon- \varepsilon_{k} -
    \frac{1}{2}\Delta- \Sigma^{R}_{11}
  \end{pmatrix},
\end{multline*}
with $\mathrm{det}(\dots)$ denoting a matrix determinant, is simply
the inverted retarded Green function to zeroth order in Poisson
brackets.  Matrix manipulations results in
\begin{multline*}
  \hat{G}^{R}(T;k,\varepsilon)
  \approx\hat{G}^{R}_{0}(T;k,\varepsilon)\\
  - \frac{\mathrm{i}\gamma\hbar^{2}}{\Delta}
  G^{R}_{0,z}(k,\varepsilon)^{2} \hat{\sigma}_{z} \partial_{T}
  (\boldsymbol{h}\cdot\hat{\boldsymbol{\sigma}}),
\end{multline*}
where we once more have used the convenient matrix notation introduced
in Eq.~\eqref{eq:4}.

A similar relation can also be found for $\hat{G}^{A}$, and we finally
use that $\hat{A} = \mathrm{i}(\hat{G}^{R} - \hat{G}^{A})$ to obtain
an expression for the spectral function linear in gradients,
\textit{viz.}
\begin{multline*}
  \hat{A}(T;k,\varepsilon)=\hat{A}_{0}(T;k,\varepsilon)\\
  +\frac{\gamma\hbar^{2}}{\Delta}\left[ G_{0,z}^{R}(k,\varepsilon)^{2}
    -G_{0,z}^{A}(k,\varepsilon)^{2}\right] \hat{\sigma}_{z}\partial
  _{T}(\boldsymbol{h}\cdot\hat{\boldsymbol{\sigma}}),
\end{multline*}
where
\begin{displaymath}
  \hat{A}_{0}(T;k,\varepsilon)=\pi\sum_{s}\delta( \varepsilon -
  \varepsilon_{ks})\left(\hat{1} + s\hat{\sigma}_{z} +
    s\frac{\gamma\hbar\boldsymbol{h}
      \cdot\hat{\boldsymbol{\sigma}}}{\Delta}\right)
\end{displaymath}
is the spectral function to zeroth order in gradients. Using that we
can rewrite
\begin{displaymath}
  G_{0,z}^{R}(k,\varepsilon)^{2} - G_{0,z}^{A}(k,\varepsilon)^{2} =
  \frac{\mathrm{i}}{\Delta}A_{z} (k,\varepsilon) +
  \frac{\mathrm{i}}{2} \partial_{\varepsilon}A_{d}(k,\varepsilon),
\end{displaymath}
we find the spectral function
\begin{multline}
  \hat{A}(T;k,\varepsilon)=\hat{A}_{0}(T;k,\varepsilon) \\
  +\frac{\mathrm{i}\gamma\hbar^{2}}{\Delta^{2}}\left(
    A_{z}(k,\varepsilon
    )+\frac{\Delta}{2}\partial_{\varepsilon}A_{d}(k,\varepsilon)\right)
  \hat{\sigma}_{z}\partial_{T} (\boldsymbol{h} \cdot
  \hat{\boldsymbol{\sigma}}).
\end{multline}
Hence, the weak, transverse driving field induces only off-diagonal
gradient contributions to the spectral function. These gradient
corrections prove to be essential in order to correctly capture the
transverse magnetization dynamics.

\end{document}